\documentclass[conference, 10pt]{IEEEtran}
\usepackage{setspace}
\usepackage{amsmath,amssymb}
\usepackage{amsthm}
\usepackage{algorithm, algpseudocode}
\usepackage{graphicx}
\usepackage{cite}

\newtheorem{claim}{Claim}[section]
\newtheorem{lemma}[claim]{Lemma}

\newtheorem{theorem}{Theorem}
\newtheorem{proposition}[claim]{Proposition}

\newtheorem{definition}[claim]{Definition}

\theoremstyle{definition}
\newtheorem{remark}[claim]{Remark}

\def\<{\langle}
\def\>{\rangle}

\def\eps{{\varepsilon}}

\def\sT{{\sf T}}

\def\E{{\mathbb E}} %expectation
\def\Var{{\sf{Var}}}

\def\reals{\mathbb{R}}

\def\normal{{\sf N}}
\def\Ber{{\sf Ber}}

\def\Var{{\sf Var}}
\def\cL{{\cal L}}

\def\d{{\mathrm{d}}}

\def\Var{{\sf Var}}

\newcommand\norm[1]{\lVert{#1}\rVert}
\newcommand\abs[1]{\left\lvert{#1}\right\rvert}

\newcommand\myeqref[1]{{Eq.\,\eqref{#1}}}

\def\sb{{\sf b}}
\def\sd{{\sf d}}

\def\hx{\widehat{x}}
\def\hu{\widehat{u}}
\def\hv{\widehat{v}}
\def\hX{\widehat{X}}
\def\Var{{\rm Var}}
\def\mse{{\sf mse}}

\def\Mmmse{{\sf M\text{-}mmse}}
\def\Smmse{{\sf S\text{-}mmse}}
\def\MSEAMP{{\sf MSE_{AMP}}}
\def\mseAMP{{\sf mse_{AMP}}}
\def\md{{\mathrm{d}}}

\author{
  \IEEEauthorblockN{Yash Deshpande}
\IEEEauthorblockA{
Department of Electrical Engineering\\
Stanford, CA. 
}
\and \IEEEauthorblockN{Andrea Montanari}
\IEEEauthorblockA{
  Departments of Electrical Engineering and Statistics \\
  Stanford, CA.
}
}
\title{Information-theoretically Optimal Sparse PCA}
\begin{document}
\maketitle

\begin{abstract}
  Sparse Principal Component Analysis (PCA) is
  a dimensionality reduction technique wherein
  one seeks a low-rank representation of a data matrix 
  with additional sparsity constraints on the 
  obtained representation. We consider two probabilistic
  formulations of sparse PCA: a spiked Wigner
  and spiked Wishart (or spiked covariance)
  model. We analyze an Approximate Message
  Passing (AMP) algorithm to estimate the
  underlying signal and show, in the high dimensional
  limit, that the AMP estimates are 
  information-theoretically optimal. As an
  immediate corollary, our results demonstrate
  that the posterior expectation of the underlying
  signal, 
  which is often intractable to compute, can be
  obtained using a polynomial-time scheme. Our results also 
  effectively provide a \emph{single-letter characterization} of the sparse PCA
  problem. 
\end{abstract}
\section{Introduction}
Suppose we are given data $Y_\lambda\in\reals^{n\times n}$ distributed according to the following \emph{spiked Wigner} model:
\begin{align}
  Y_{\lambda} &= \sqrt\frac{\lambda}{n}xx^\sT + Z \label{eqn:modelWig}.
\end{align}
Here $x \in\reals^n$, and each coordinate $x_i$ is an independent Bernoulli random
variable with probability $\eps$, denoted by $x_i \sim\Ber(\eps)$.  $Z\in\reals^{n\times n}$ is a symmetric
matrix where $(Z_{ij})_{i\le j}$ are i.i.d $\normal(0, 1)$ variables, independent of $x$.  
Analogously, consider the following \emph{spiked Wishart}
model:
\begin{align}
  Y_\lambda &= \sqrt\frac{\lambda}{n}uv^\sT + Z \label{eqn:modelWish}.
\end{align}
Here $u\in\reals^m$, with i.i.d coordinates $u_i\sim\normal(0, 1)$ and 
$v\in\reals^n$ with i.i.d Bernoulli coordinates $v_j \sim\Ber(\eps)$. Further, $Z\in\reals^{m\times n}$ is
a matrix with $Z_{ij}\sim\normal(0, 1)$ i.i.d. random variables.

In either case, our data consists 
of a \emph{sparse, rank-one} matrix observed through Gaussian noise. We let
$X$ denote the clean, underlying signal ($xx^\sT$ or $uv^\sT$ for the
spiked Wigner or Wishart model respectively). Our task is to estimate
the signal $X$ from the data $Y_\lambda$ in the high dimension asymptotic
where $n\to\infty, m\to\infty$ with $m/n \to\alpha \in (0, \infty)$. 
This paper focuses on estimation in the sense of the mean
squared error, defined for an estimator $\hX(Y_\lambda)$ as:
\begin{align}\label{eqn:mseXdef}
  \mse(\hX, \lambda) &\equiv \frac{1}{n^2}\E\left\{ \norm{\hX - X}_F^2 \right\}. 
\end{align}

It is well-known \cite{cover2012elements} that the mean squared error is minimized by the estimator $\hX = \E\{X|Y_\lambda\}$, 
i.e. the conditional expectation of the signal given the observations. Consequently, 
the minimum mean squared error (MMSE) is given by:
\begin{align}
  \Mmmse(\lambda, n) &\equiv \frac{1}{n^2}\E\left\{ \norm{X - \E\{X|Y_\lambda\}}_F^2 \right\}. \label{eqn:coordmmse}
\end{align}
In this paper, we analyze an iterative scheme called \emph{approximate message passing} (AMP)
to estimate the clean signal $X$. 
The machinery of approximate message passing reduces the
high-dimensional \emph{matrix} problem in models
\eqref{eqn:modelWig}, \eqref{eqn:modelWish} to the following simpler
\emph{scalar} denoising problem:
\begin{align}
  Y_\lambda &= \sqrt{\lambda}X_0 + N, \label{eqn:scalarDenoise}
\end{align}
where $X_0\sim\Ber(\eps)$ and $N\sim\normal(0, 1)$ are independent.
The scalar MMSE\cite{guo2005mutual} in estimating $X_0$ from $Y_\lambda$ is given by:
\begin{align*}
  \Smmse(X_0, \lambda) = \E\left\{ (X_0 - \E\{X_0|Y_\lambda\})^2 \right\}.
\end{align*}
Our main results, 
characterize the
optimal mean squared error $\Mmmse(\lambda, n)$ in the large $n$ asymptotic, when $\eps>\eps_c\approx 0.05$,
and establish that AMP achieves this fundamental limit. 
For the spiked Wigner model we prove the following.
\begin{theorem}
  There exists an $\eps_c \in (0, 1)$ such that for all $\eps>\eps_c$,
  and every $\lambda \ge 0$ the squared error of AMP iterates $\hX^t$ satisfies the following:
  \begin{align*}
    \lim_{t\to\infty}\lim_{n\to\infty}\mse(\hX^t, \lambda) &= \lim_{n\to\infty} \Mmmse(\lambda, n).
  \end{align*}
  Further, the limit on the RHS above satisfies, for every $\lambda>0$:
  \begin{align*}
    \lim_{n\to\infty}\Mmmse(\lambda, n) &= \eps^2 - y_*^2/\lambda^2,
  \end{align*}
  where $y_* =y_*(\lambda)$ solves $y_* = \lambda(\eps - \Smmse(X_0, y_*))$. 
  \label{thm:introthm}
\end{theorem}
Some remarks are in order:
\begin{remark}
  The combination of Theorem \ref{thm:introthm} and \myeqref{eqn:fixedptWig} effectively yield a \emph{single-letter characterization}
  of Model \eqref{eqn:modelWig}, connecting the limiting matrix MMSE with the MMSE
  of a calibrated scalar denoising problem $\Smmse(X_0, y_*(\lambda))$. 
\end{remark}
\begin{remark}
  It is straightforward to establish that $\eps_c<1$. However, numerically we obtain that 
  $\eps_c \approx 0.05$. Thus, for most values of $\eps$, 
  our results completely characterize the spiked Wigner model. 
\end{remark}
\subsection*{Background and Motivation}
Probabilistic models similar to Eqs.\eqref{eqn:modelWig}, \eqref{eqn:modelWish}
have been the focus of much recent work in random matrix theory \cite{furedi1981eigenvalues, knowles2013isotropic, benaych2011eigenvalues, baik2005phase, baik2006eigenvalues}. 
The focus
in this literature is to analyze the limiting distribution of the eigenvalues 
of the matrix $Y_\lambda/\sqrt{n}$ and, in particular, 
identifying regimes in which this distribution differs from that of the pure noise
matrix $Z$. The typical picture that emerges from this line of work is that a
\emph{phase transition} occurs at a well-defined critical signal-to-noise ratio $\lambda_c = \lambda_c(\alpha, \eps)$:
\begin{LaTeXdescription}
  \item[Above the threshold ${\lambda > \lambda_c}$:] there exists an outlier eigenvalue and
    the principal eigenvector corresponding to this outlier has
    a positive correlation with the signal. For instance, in the spiked Wigner case, 
    letting $\hx_1(Y_\lambda)$ denote the normalized principal eigenvector of $Y_\lambda$ we 
    obtain $\<\hx_1(Y_\lambda), x\>/\sqrt{n\eps} \ge \delta(\eps) >0$ asymptotically. 
  \item[Below the threshold $\lambda < \lambda_c$:] the spectral distribution of
    the observation $Y_\lambda$ is indistinguishable from that of the pure noise
    $Z$. Furthermore, the principal eigenvector  is asymptotically
    \emph{orthogonal} to the signal factors. For the spiked Wigner case, this
    implies  that $\<\hx_1(Y_\lambda), x\>/\sqrt{n\eps} \to 0$ asymptotically. 
\end{LaTeXdescription}
This phase transition phenomenon has been demonstrated under considerably fewer
assumptions than we make in Eqs. \eqref{eqn:modelWig}, \eqref{eqn:modelWish}. We
refer the interested reader to \cite{pizzo2013finite, knowles2013isotropic} and the references therein for 
further details. 

It is clear
from these results that vanilla PCA, which involves using the principal 
eigenvector is ineffective in estimating the 
underlying clean signal $X$ when $\lambda<\lambda_c$. Indeed PCA
only makes use of the fact that the underlying signal is \emph{low-rank}, or
in fact rank-one in our case. Since we make additional sparsity assumptions 
in our models \eqref{eqn:modelWig}, \eqref{eqn:modelWish} it is natural to ask if 
this can be leveraged when we have a small signal-to-noise ratio $\lambda$. 
In the last decade, a considerable amount of work in the statistics community has studied 
this problem.  Our spiked Wishart model \myeqref{eqn:modelWish} is a special case
of the \emph{spiked covariance model}
in statistics, first introduced by Johnstone
and Lu \cite{johnstone2004sparse, johnstone2009consistency}. 
Johnstone and Lu proposed a simple diagonal thresholding
scheme that estimates the support of $v$ using the largest diagonal
entries of the Gram matrix $Y_\lambda^\sT Y_\lambda$. An M-estimator
for the underlying factors was proposed by \cite{d2007direct}. A number of other
practical algorithms \cite{zou2006sparse, moghaddam2005spectral, d2008optimal} have also been proposed to outperform diagonal
thresholding. 

Some recent work \cite{KrauthgamerSPCA, deshpande2013sparse} has focused on the \emph{support recovery}
guarantees for such algorithms, or estimating consistently the positions of non-zeros in $v$. 
Let $k = n\eps$ denote the expected size of the support of $v$. 
Amini and Wainwright \cite{amini2009high} proved that unless $k \le c m/\log n$, 
\emph{no algorithm} would be able to consistently estimate the support of $v$
due to information-theoretic obstructions. They further demonstrate that
a (computationally intractable) algorithm that searches through all possible 
$k$-sized subsets of rows of
the data matrix can recover the support provided $k \le c' m/\log n$. 

Since we consider $\eps = \Theta(1)$ and $m = \Theta(n)$, in our case $k = \Theta(m)$ and
consequently, estimating the support correctly is impossible. It is
for this reason that we instead
focus on another natural figure-of-merit: the mean squared error, defined
in \myeqref{eqn:mseXdef} above. 
Somewhat surprisingly, we are able to prove (for a regime $\eps_c < \eps \le 1$) 
that a \emph{computationally efficient}
algorithm asymptotically achieves the information-theoretically optimal mean squared error \emph{for 
any signal-to-noise ratio $\lambda$}.
%\vspace{-2pt}
%coordinate matrix minimum mean squared error 
%function as:
%The I-MMSE identity of \cite{} further allows to connect $\mmse(\lambda, n)$ with 
%the mutual information $I(X; Y_\lambda)$. As we will see, 
%approximate message passing allows us to view the high-dimensional \emph{vector} or \emph{matrix} problem of
%Eqs.\eqref{eqn:modelWig}, \eqref{eqn:modelWish} in terms of a much 
%simpler \emph{scalar} denoising problem:
%\begin{align*}
%  Y_\lambda &= \sqrt{\lambda}X_0 + Z,
%\end{align*}
%where $X_0\sim\Ber(\eps)$ and $Z\sim\normal(0, 1)$ are independent. The \emph{scalar}
%MMSE function is defined as:
%\begin{align*}
%  \Smmse(X_0, \lambda) &\equiv \E\left\{ (X_0 - \E\{X_0 | Y_\lambda\})^2 \right\}.
%\end{align*}
%Analogous functions can be defined for different random variables, hence 
%we retain the dependence on $X_0$ in our notation.
%The properties of $\Smmse(X_0, \lambda)$ then allow us to relate the performance
%of AMP with the normalized mutual information $I(X;Y_\lambda)/n$, in the high-dimensional limit
%of $n\to\infty$. In conjunction with the I-MMSE identity, this allows us to show that,
%asymptotically in $n$, AMP achieves the optimal mean squared error $\Mmmse(\lambda)=\lim_{n\to\infty}\Mmmse(\lambda, n)$.
\subsection*{Other related work}
Rangan et al. \cite{rangan2012iterative} considered a model similar to
\myeqref{eqn:modelWish} with general structural assumptions
on the factors $u$ and $v$. They proposed an approximate
message passing algorithm analogous to the one we analyze
and characterize its high-dimensional behavior. Based on non-rigorous
but powerful tools from statistical physics, they conjecture that AMP 
asymptotically achieves the (optimal) 
performance of the joint
MMSE estimator of $u$ and $v$.  In the restricted setting
of sparse PCA, we rigorously confirm this conjecture, and validate
the statistical physics arguments. 

A model similar to \myeqref{eqn:modelWig} was considered by \cite{deshpande2013finding},
motivated by the ``planted clique'' problem in theoretical
computer science. The sparsity regime of interest in this work was 
$k = O(\sqrt{n})$, with a focus on recovering the ``clique'', 
analogous to support recovery in the spiked covariance model. 
%\vspace{-0.5cm}
\subsection*{Organization}
The paper is organized as follows. In Section \ref{sec:alg} we give details of the AMP
algorithm and formally state our results. For brevity, we only provide
the proof of one of our main results in Section \ref{sec:proof}. 
\section{Algorithm and main results}\label{sec:alg}
In the interest of exposition, we restrict ourselves to the
spiked Wigner model \eqref{eqn:modelWig}  and defer the discussion of the Wishart model
\eqref{eqn:modelWish} to Section \ref{subsec:Wishart}.
%\vspace{-5pt}
\subsection{Approximate Message Passing}
Approximate message passing (AMP) is a low complexity iterative
algorithm that produces iterates $x^t, \hx^t\in\reals^n$ 
For a data matrix $A$ we define for $t \ge 0$:
\begin{IEEEeqnarray}{rCl}
  x^{t+1} &=& A\hx^t - \sb_t \hx^{t-1} \\
  \hx^t &=& f_t(x^t).
  \label{eqn:ampIterDef}
\end{IEEEeqnarray}
Here $f_t:\reals\to\reals$ are scalar functions and $\{\sb_t\}_{t\ge 0}$ is a sequence of
scalars. Here and below, for a scalar function $f$, we define its extension to 
$\reals^n$ by applying it component-wise, i.e. 
$f:\reals^n \to\reals^n,  v \mapsto f(v) = (f(v_1), f(v_2) \cdots , f(v_n))^{\sT}$.
We further define the matrix estimate $\hX^t \equiv \hx^t(\hx^t)^\sT$. For the complete
description of the algorithm, we refer the reader to Algorithm \ref{alg:ampsym} below, which provides prescriptions
for the functions $f_t$ and the scalars $\sb_t$. 

\begin{algorithm}
  \caption{Symmetric Bayes-optimal AMP}
  \label{alg:ampsym}
    {\bf Input:} Data $Y_\lambda$ as in \myeqref{eqn:modelWig}\\
   Define $A = Y_\lambda/\sqrt{n}$, and $\hx^{0}, \hx^{-1} = 0$. For $t\ge 0$ compute
    \begin{align*}
      x^{t+1} &= A\hx^t - \sb_t\hx^{t-1}\\
      \hx^t &= f_t(x^t) \\
      \hX^t &= \hx^t(\hx^t)^\sT,
    \end{align*}
    where $f_t(y):\reals\to\reals$ is recursively defined:
    \begin{align*}
      f_t(y) &= \E\{X_0 | \mu_tX_0 + \sqrt{\tau_t}Z = y\}.
    \end{align*}
    Here $X_0\sim\Ber(\eps)$ and $Z\sim\normal(0, 1)$ are independent, and
    $\mu_t, \tau_t$ are defined as in Eqs. \eqref{eqn:SEsym1}, \eqref{eqn:SEsym2}. 
    Also the scalars $\sb_t$ are computed as:
    \begin{align*}
      \sb_t &= \frac{1}{n}\sum_{i=1}^nf_t'(x^t_i).
    \end{align*}
    %\vspace{-0.1cm}
\end{algorithm}
%\begin{algorithm} 
%  \caption{Asymmetric Bayes-optimal AMP}
%  \label{alg:ampasym}
%    {\bf Input:} Data $Y_\lambda$ as in \myeqref{eqn:modelWig}\\
%   Define $A = Y_\lambda/\sqrt{n}$, and $\hu^{0}, \hu^{-1} = 0$. For $t\ge 0$ compute
%    \begin{align*}
%      \tu^{t+1} &= A\hv^t - \sb_t\hu^{}\\
%      \hu^t &= f_t(\tu^t) \\
%      \tv^t &= A^\sT\hu^t - \sd_t\hv^{t-1} \\
%      \hu^t &= g_t(\tu^t) \\
%      \hX^t &= \hu^t(\hv^t)^{\sT},
%    \end{align*}
%    where $f_t(y):\reals\to\reals$ and $g_t(y):\reals\to\reals$ are defined by
%    \begin{align*}
%      f_t(y) &= \E\{V | \mu^v_tV + \sqrt{\tau^v_t}Z = y\}\\
%      g_t(y) &= \E\{U | \mu^u_tU + \sqrt{\tau^u_t}Z = y\}.
%    \end{align*}
%    Here $V\sim\Ber(\eps)$, $U, Z\sim\normal(0, 1)$ are independent and
%    $\mu^u_t, \tau^u_t, \mu^v_t, \tau^v_t$ are defined as in Eqs. \eqref{eqn:SEasym1}, \eqref{eqn:SEasym2}, \eqref{eqn:SEasym3} and \eqref{eqn:SEasym4}. 
%    Also the scalars $\sb_t, \sd_t$ are computed as:
%    \begin{align*}
%      \sb_t &= \frac{1}{n}\sum_{i=1}^nf_t'(\tv^t_i).
%    \end{align*}
%    \begin{align*}
%      \sd_t &= \frac{1}{n}\sum_{i=1}^ng_t'(\tu^t_i).
%    \end{align*}
%    {\bf Y: CONFIRM ONSAGER}
%\end{algorithm}
%\vspace{-10pt}
\subsection{State evolution}
The key property of approximate message passing
is that it admits an \emph{asymptotically exact}
characterization in the high-dimensional limit where $n\to\infty$.
The iterates $x^t_i$  converge as $n\to\infty$
to Gaussian random variables with a prescribed mean and variance. 
These prescribed mean and variance parameters evolve according to
\emph{deterministic} recursions, jointly termed ``state evolution''. 
We define for $t\ge 0$:
\begin{IEEEeqnarray}{rCl}
  \mu_{t+1} &=& \sqrt{\lambda}\E\{X_0f_t(\mu_t X_0 + \sqrt{\tau_t} Z)\}\label{eqn:SEsym1} \\
  \tau_{t+1} &=& \E\{f_t(\mu_tX_0 + \sqrt{\tau_t} Z)^2\},
  \label{eqn:SEsym2}
\end{IEEEeqnarray}
where $X_0 \sim \Ber(\eps)$ and $Z\sim\normal(0, 1)$ are independent. The
recursion is initialized with $\mu_0 = \tau_0 = 0$. 

The state evolution recursions succinctly describe the iterates arising
in AMP. Formally, we have, for any continuous function
$\psi:\reals^2\to\reals$ the following is true wherever the expectation on
the right is defined:
\begin{align*}
  \lim_{n\to\infty} \frac{1}{n}\sum_{i\in n}\psi(x_i, x^t_i) = \E\{\psi(X_0, \mu_tX_0 +\sqrt{\tau_t}Z)\} \text{ a.s. }
\end{align*}
where $\mu_t, \tau_t$ are defined by Eqs. \eqref{eqn:SEsym1}, \eqref{eqn:SEsym2}. 
This allows us to track the squared error of the 
AMP estimator accurately, in the high-dimensional limit, and establish
its optimality.

Although we define AMP and the corresponding
state evolution for general scalar functions $f_t$, our prescription 
Algorithm \ref{alg:ampsym} uses specific choices for $f_t$. 
In the spiked Wigner case, we choose $f_t(y) = \E\{X_0|\mu_tX_0 + \sqrt{\tau_t}Z = y\}$, 
the posterior expectation of $X_0$, with observation corrupted by Gaussian 
noise and SNR $\mu_t^2/\tau_t$. 
To stress this fact, we will refer to our algorithms as Bayes-optimal AMP.

\subsection{Main Result}

\begin{figure}[th!] 
  \centering
  \includegraphics[width=8cm]{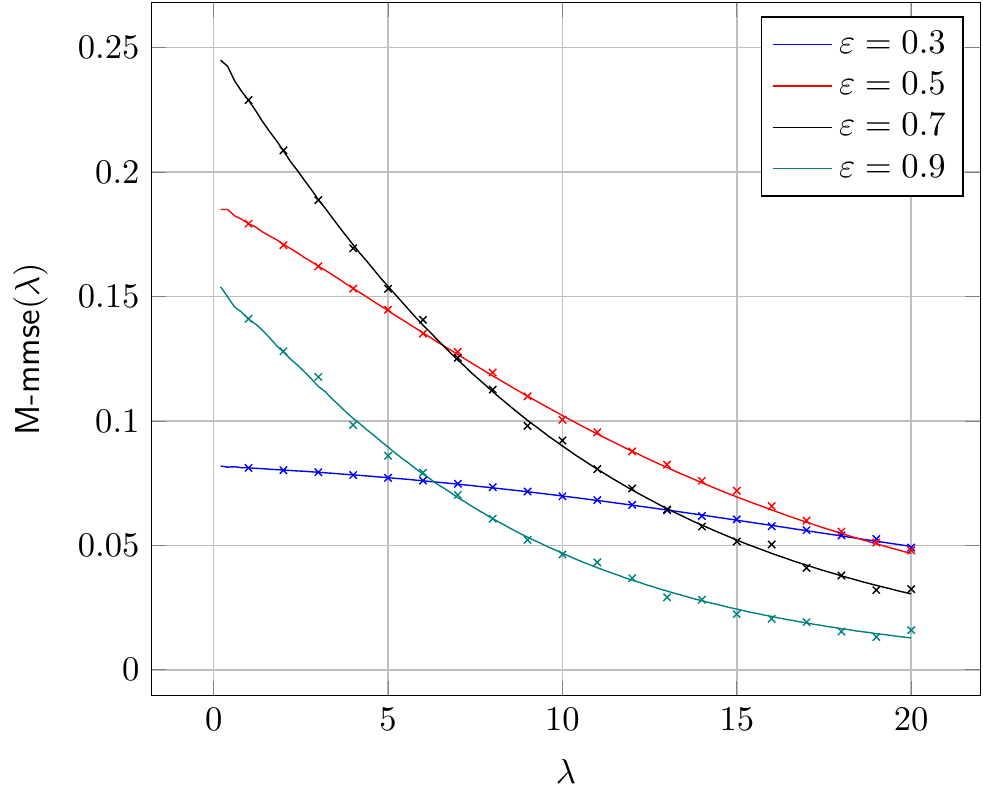}
  \caption{The solid curves $\Mmmse(\lambda)$ above are computed analytically using Theorem \ref{thm:main}.
  The crosses mark median MSE incurred by AMP in $100$ Monte Carlo runs
  with $n=2000$ for the spiked Wigner model \eqref{eqn:modelWig}.\label{fig:mseAMP}}
  %\vspace{-2cm}
\end{figure}

We first define the following regime for $\eps$:

\begin{definition}\label{def:Wig}
Let $\eps_*\in(0, 1)$ be the smallest positive real number such 
that for every $\eps>\eps_*$ the following is true. For every $\lambda>0$, 
the equation below has \emph{only one} solution in $[0, \infty)$:
  \begin{align}
    \lambda^{-1} y = \eps - \Smmse(X_0, y).\label{eqn:fixedptWig}
  \end{align}
  Here $X_0 \sim\Ber(\eps)$.
\end{definition}
With a slight abuse of notation, we denote by $\Mmmse(\lambda)$ the
quantity $\lim_{n\to\infty}\Mmmse(\lambda, n)$, assuming
it exists. Also we define the squared error of AMP at iteration
$t$ as:
\begin{align*}
  \MSEAMP(\lambda, t) &= \frac{1}{n^2}\norm{\hX^t - X}_F^2.
\end{align*}
Notice that $\MSEAMP(\lambda, t)$ is a random variable that depends
on the realization $Y_\lambda$. Our first main result strengthens Theorem 
\ref{thm:introthm} for the spiked
Wigner case:
\begin{theorem}  \label{thm:main}
  Under Model \eqref{eqn:modelWig} we have
  $\Mmmse(\lambda) = \lim_{n\to\infty}\Mmmse(\lambda, n)$ exists for every $\lambda\ge 0$. This
  limit satisfies, when $\eps>\eps_*$:
  \begin{align}
    \Mmmse(\lambda) &= \eps^2 - \frac{y_*(\lambda)^2}{\lambda^2}, \label{eqn:mainClaim1}
  \end{align}
  where $y_*(\lambda)$ is the unique solution to \myeqref{eqn:fixedptWig} above. 
  Further, the symmetric Bayes-optimal AMP algorithm \ref{alg:ampsym} satisfies the following 
  almost surely:
      \begin{align}
	\lim_{t\to\infty}\lim_{n\to\infty}\MSEAMP(\lambda, t) = \Mmmse(\lambda). \label{eqn:mainClaim2}
      \end{align}
\end{theorem}
Although this result is asymptotic in nature, simulations show that the
predictions are accurate on problems of dimension a few thousands (see
Figure \ref{fig:mseAMP}).
\subsection{The spiked Wishart model} \label{subsec:Wishart}
An asymmetric version of Algorithm \ref{alg:ampsym}
can also be written. It involves iterates $u^t, \hu^t\in\reals^m$, 
$v^t, \hv^t\in\reals^n$. Define $A = Y_\lambda/\sqrt{m}$, and 
$\hu^{0}, \hu^{-1} = 0$. For $t\ge 0$ compute
    \begin{align*}
      u^{t+1} &= A\hv^t - \sb_t\hu^{t}\\
      \hv^t &= f_t(v^t) \\
      v^t &= A^\sT\hu^t - \sd_t\hv^{t-1} \\
      \hu^t &= g_t(u^t) \\
      \hX^t &= \hu^t(\hv^t)^{\sT},
    \end{align*}
    The following is the analogue of Definition \ref{def:Wig} for the
    asymmetric Wishart model:
    \begin{definition}\label{def:Wish}
Let $\widetilde{\eps}_*$ be the smallest positive real number such that
for every $\eps>\widetilde{\eps}_*$ the following is true. For every 
$\lambda>0$ equation below has \emph{only one} solution in $[0, \infty)$:
  \begin{align}
    \lambda^{-1} y = \eps - \Smmse(V, \lambda\alpha y/(1+y))\label{eqn:fixedptWish}.
  \end{align}
  Here $V \sim\Ber(\eps)$.
\end{definition}
Our second result is for the spiked Wishart model \eqref{eqn:modelWish}:
\begin{theorem} \label{thm:main2}
  The limit $\Mmmse(\lambda) \equiv \lim_{n\to\infty}\Mmmse(\lambda, n)$ exists for every $\lambda\ge 0$
  and, for $\eps>\widetilde{\eps}_*$, is given by:
  \begin{align}
    \Mmmse(\lambda) &= \eps - \frac{y_*(\lambda)^2}{\lambda(1+y_*(\lambda))},\label{eqn:mainClaim3}
  \end{align}
  where $y_*(\lambda)$ is the unique solution to \myeqref{eqn:fixedptWish}.
    Further,   asymmetric Bayes-optimal AMP satisfies the following limits almost surely:
      \begin{align}
	\lim_{t\to\infty}\lim_{n\to\infty}\MSEAMP(\lambda, t) = \Mmmse(\lambda). \label{eqn:mainClaim4}
      \end{align}
\end{theorem}
\begin{remark}
  Numerically verifying the fixed point condition \myeqref{eqn:fixedptWish} is somewhat more
  involved than \myeqref{eqn:fixedptWig}. However it is relatively straightforward to 
  establish that $\widetilde{\eps}_* \le \inf\{\eps: \Smmse(V, z) \text{ is convex in } z\} \approx 0.05$.   
\end{remark}

%\begin{figure}[tbp]
%  \centering
%  \includegraphics[width=8cm]{./FIGS/mmse.pdf}
%  \caption{The scalar MMSE curves $\Smmse(\lambda)$ for different values of $\eps$. The curves for
%    $\eps=0.3$ and $\eps=0,7$ coincide. Numerically, 
%    we see that these satisfy the condition of Definition \ref{def:Wig}.}
%  \label{fig:mmse}
%  \vspace{-0.5cm}
%\end{figure}
\section{Proof of Theorem \ref{thm:main}}\label{sec:proof}

Owing to space constraints, we 
restrict ourselves to proving Theorem \ref{thm:main} in this paper. 
The proof of Theorem \ref{thm:main2} follows similar ideas and
will be provided in the full version of the present paper.
Theorem \ref{thm:main} follows almost immediately from the following two propositions.

\begin{proposition}\label{prop:ampmse}
  Consider the model \myeqref{eqn:modelWig} with $\eps > \eps_*$, and the approximate message 
  passing orbit obtained by using the recursively defined scalar functions for $t\ge 0$:
  \begin{IEEEeqnarray*}{rCl}
    f_t(y) &=& \E\{X_0 | \mu_{t}X_0 + \sqrt{\tau_t} Z=y\}.
  \end{IEEEeqnarray*}
  Here $X_0\sim\Ber(\eps)$ and $Z\sim\normal(0, 1)$ are independent. Further $\mu_0 = \tau_0 = 0$ and
  $(\mu_t, \tau_t)_{t\ge 1}$ are defined using the state evolution recursions \eqref{eqn:SEsym1}, \eqref{eqn:SEsym2}.
  Then defining $\mseAMP(\lambda)\equiv \eps^2 - y_*(\lambda)^2/\lambda^2$, the RHS of \myeqref{eqn:mainClaim1}, the following is true:
  \begin{IEEEeqnarray}{rCl}
    \lim_{t\to\infty}\lim_{n\to\infty}\MSEAMP(\lambda, t) &=& \mseAMP(\lambda) \label{eqn:claim1}\\
    \int_0^{\infty} \mseAMP(\lambda)\mathrm{d}\lambda &=& 4h(\eps)\label{eqn:claim2}.
  \end{IEEEeqnarray}
  The first limit holds almost surely and in $\cL_1$ and 
  $h(\eps)$ is the binary entropy function $h(\eps) = -\eps\log\eps - (1-\eps)\log(1-\eps)$.
\end{proposition}

\begin{proposition}\label{prop:immse}
  For every $\lambda\ge 0$, $\Mmmse(\lambda) = \lim_{n\to\infty}\Mmmse(\lambda, n)$ exists. Further:
  \begin{IEEEeqnarray}{rCl}
    \int_{0}^\infty  \Mmmse(\lambda)\md\lambda &\ge& 4h(\eps), \label{eqn:immseclaim}
  \end{IEEEeqnarray}
  where $h(\eps)$ is defined in Proposition \ref{prop:ampmse}.
\end{proposition}

The above propositions are proved in Subsections
\ref{prop:ampmse} and \ref{prop:immse} respectively. We
first use these to establish Theorem \ref{thm:main}.
%\begin{proof}[Proof of Theorem \ref{thm:main}]
Since the posterior expectation minimizes the mean
squared error, we have that $\Mmmse(\lambda, n) \le \E\{\MSEAMP(\lambda, t)\}$. 
Taking the limits $n\to\infty, t\to\infty$ in that order,
and employing the first claim of Proposition \ref{prop:ampmse} 
we have that:
\begin{align*}
  \Mmmse(\lambda) &\le \mseAMP(\lambda).
\end{align*}
This implies that:
\begin{align*}
  4h(\eps) &\le \int_{0}^{\infty}\Mmmse(\lambda)\d\lambda \\
  &\le \int_0^\infty \mseAMP(\lambda)\d\lambda = 4h(\eps),
\end{align*}
where in the first inequality and the last equality we use
Propositions \ref{prop:ampmse}, \ref{prop:immse}. This implies
that $\mseAMP(\lambda) = \Mmmse(\lambda)$ for Lebesgue-a.e. $\lambda$. 
Further, as $\Mmmse(\lambda)$ is the pointwise limit of monotone non-increasing
(in $\lambda$) functions $\Mmmse(\lambda, n)$\cite{guo2011estimation}, it is monotone non-increasing,
which yields the claim for all $\lambda\in [0, \infty)$. \qedsymbol
%\end{proof}
\subsection{Proof of Proposition \ref{prop:ampmse}}
Note that:
\begin{align*}
  \MSEAMP(\lambda, t) &= \frac{1}{n^2}\norm{\hX^t - X}_F^2 \\
  &= \frac{1}{n^2}(\norm{\hx^t}^4 + \norm{x}^4 -2\<\hx^t, x\>^2).
\end{align*}
By the strong law of large numbers, $\norm{x}^4/n^2 \to \eps^2$ 
almost surely, and in $\cL_1$. It is not hard to prove
that the functions $f_t(y)$ are $\sqrt{\lambda}$-Lipschitz
continuous. Hence, it is a direct consequence of Theorem 1 of \cite{javanmard2012state}
that the following limits hold almost surely and in $\cL_1$:
\begin{align*}
  \lim_{n\to\infty} \frac{1}{n^2}\norm{\hx^t}^4 &= (\E\{f_t(\mu_t X_0 + \sqrt{\tau_t}Z)^2\})^2 \\
  \lim_{n\to\infty}\frac{1}{n^2}\<\hx^t, x\>^2 &= (\E\{ X_0f_t(\mu_tX_0 + \sqrt{\tau_t}Z)  \} )^2.
\end{align*}

Further, our choice $f_t(y) = \E\left\{ X_0\vert \mu_tX_0+ \sqrt{\tau_t}Z = y\right\}$ yields, 
by use of the tower property of conditional expectation:
\begin{align*}
  \E\left\{ X_0f_t(\mu_tX_0 + \sqrt{\tau_t}Z )\right\} &= \E\left\{ f_t(\mu_tX_0 + \sqrt{\tau_t}Z)^2\right\} \\
  &= \tau_{t+1}.
\end{align*}
It follows that $\lim_{t\to\infty}\lim_{n\to\infty}\MSEAMP(\lambda, t) = \eps^2 - \tau_*^2$
almost surely and in $\cL_1$ where $\tau_* = \tau_*(\lambda)$ denotes the smallest non-negative fixed point
of the equation:
\begin{align}
  \tau &= \E\left\{\E\{ X_0 \vert \sqrt{\lambda}\tau X_0 + \sqrt{\tau}Z \}^2 \right\}.\label{eqn:fixedptdef}
\end{align}
Since the right hand side equals $\eps(1-\eps)$ at $\tau = 0$ and $\eps$
at $\tau=\infty$, at least one fixed point must exist. Hence $\tau_*(\lambda)$
is well defined. Now, note that
\begin{align*}
  \E\left\{\E\{ X_0 \vert \sqrt{\lambda}\tau X_0 + \sqrt{\tau}Z \}^2 \right\} &= \E\left\{\E\{ X_0 \vert \sqrt{\lambda\tau}X_0 + Z \}^2 \right\} \\
  &= \E\{X_0^2\} - \Smmse(X_0, {\lambda\tau}) \\
  &= \eps - \Smmse(X_0, {\lambda\tau}).
\end{align*}
Thus $\tau_*$ is a fixed point of \myeqref{eqn:fixedptdef} iff $\lambda\tau_*=y_*$ is
a fixed point of \myeqref{eqn:fixedptWig}. It follows from our definition of $\eps_*$
that when $\eps>\eps_*$, $\tau_*(\lambda)$ is the unique non-negative fixed point of
\myeqref{eqn:fixedptdef} and Claim \eqref{eqn:claim1} follows. 
  To complete the proof of the proposition, it only remains
to show claim \eqref{eqn:claim2}, for which we have the 
following 
\begin{lemma} 
  \label{lem:replica}
  Let $\tau_*(\lambda)$ denote the unique non-negative fixed point of \myeqref{eqn:fixedptdef}. 
  Then 
  \begin{align*}
    \int_{0}^{\infty}(\eps^2 - \tau_*(\lambda))^2 \d\lambda &= 4h(\eps).
  \end{align*}
\end{lemma}
\begin{proof}
  Define the function:
  \begin{align*}
    \phi(\lambda, m) &= \frac{m^2}{4} + \frac{\eps^2\lambda^2}{4} \\
    &\quad - \E\log\left( 1-\eps + \eps\exp\{W(m, \lambda, X_0, Z)\} \right).
  \end{align*}
  where $W(m, \lambda, x, z) = m\sqrt{\lambda}x - m\sqrt{\lambda}/2 + \lambda^{1/4}m^{1/2}z$. Here
  $X_0\sim\Ber(\eps)$, $Z\sim\normal(0, 1)$ and are independent. Letting $m_* = \tau_*\sqrt{\lambda}$, 
  it is not hard to show that:
  \begin{align*}
    \frac{\partial\phi}{\partial m}\bigg\vert_{m = m_*} &= 0, \\
    \frac{\partial\phi}{\partial\lambda}\bigg\vert_{m = m_*} &=\frac{1}{4}\left(\eps^2 - \frac{m_*^2}{\lambda}\right)
  \end{align*}
  It follows from the fundamental theorem of calculus that 
  \begin{align*}
    \int_{0}^{\infty}(\eps^2 - \tau_*^2)\d\lambda &= \phi(\lambda, m_*(\lambda))\big\vert^\infty_0.
  \end{align*}
  It is easy to see that $\phi(0, m_*(0)) = 0$. Further,
  using the fact that $\tau_*(\lambda) \le 1$ (as the
  right hand side of \myeqref{eqn:fixedptdef} is bounded by 1), 
  we have that $m_*(\lambda) = O(\sqrt{\lambda})$. Using this, 
  it is not hard to check that $\phi(\lambda, m_*(\lambda)) \to h(\eps)$
  as $\lambda\to\infty$. This concludes the proof of the lemma.
\end{proof}
\subsection{Proof of Proposition \ref{prop:immse}}
We first prove that $\lim_{n\to\infty}\Mmmse(\lambda, n)$ exists
for every $\lambda\ge 0$. Define for $i, j \in [n]$
\begin{IEEEeqnarray*}{rCl}
  m_{ij}(\lambda, n) &\equiv& \E\left\{ (X_{ij} - \E\{X_{ij}\vert Y_\lambda\})^2 \right\}.
\end{IEEEeqnarray*}
By the fact that the distribution of $(X, Y_\lambda)$ is invariant
under (identical) row and column permutations, $m_{ij}(\lambda, n) = m_{12}(\lambda, n)$
for every $i, j$ distinct. Consequently:
\begin{IEEEeqnarray*}{rCl}
  \abs{\Mmmse(\lambda, n)- m_{12}(\lambda, n)} &=& \frac{1}{n}m_{11}(n, \lambda) \le \frac{\Var(X_{11})}{n}.\\
\end{IEEEeqnarray*}
Since $\Var(X_{11}) = \eps(1-\eps) <\infty$ it suffices to prove
that $\lim_{n\to\infty}m_{12}(\lambda, n)$ exists for every $\lambda\ge 0$.
To this end, let $Y_\lambda^{n-1}$ denote the first principal $(n-1)\times(n-1)$
submatrix of $Y_\lambda$. Clearly:
\begin{align*}
  m_{12}(\lambda, n) &\le \E\left\{ (X_{12} - \E\{X_{12}\vert Y_{\lambda}^{n-1}\})^2 \right\} \\
  &= m_{12}((n-1)\lambda/n, n-1) \\
  &\le m_{12}(\lambda, n-1),
\end{align*}
where the equality follows from the model \myeqref{eqn:modelWig} and the 
second inequality from monotonicity of the minimum mean square
error in $\lambda$ \cite{guo2011estimation}. Consequently, for every $\lambda\ge 0$,  $m_{12}(\lambda, n)$ 
is a monotone, bounded sequence and has a limit. 

In order to prove the claim \ref{eqn:immseclaim}, we first note that for
any finite $n$ the following holds applying the I-MMSE
identity of \cite{guo2005mutual} to the upper triangular portion
of $X$:
\begin{align*}
  I(X; Y_\Lambda) &= \frac{1}{2n} \int_{0}^{\Lambda}\left(\frac{n(n-1)}{2}m_{12}(\lambda, n) + nm_{11}(\lambda, n)\right) \d\lambda.
\end{align*}
For $\Lambda=\infty$ we have that $I(X; Y_{\infty}) = H(X) - H(X|Y_{\infty})= H(X)$ and $H(X) = H(x) = nh(\eps)$.
Dividing by $n$ on either side:
\begin{align*}
  h(\eps) &=\frac{1}{2n^2} \int_{0}^{\infty} \left(\frac{n(n-1)}{2}m_{12}(\lambda, n) + nm_{11}(\lambda, n)\right) \d\lambda.
\end{align*}
An application of Fatou's lemma then yields the result. \qedsymbol
\bibliographystyle{IEEEtran}
\bibliography{all-bibliography}
\end{document}